\documentclass[conference]{IEEEtran}
\IEEEoverridecommandlockouts

\usepackage[utf8]{inputenc}
\usepackage[T1]{fontenc}
\usepackage{microtype,newtxtext,newtxmath} 
\usepackage{mathtools,gensymb,url} 
\usepackage{subcaption}%
\usepackage[font=footnotesize,textfont=footnotesize]{subcaption}
\usepackage{longtable,booktabs,array}
\usepackage[backend=biber, sorting=none, style=ieee, maxnames=5]{biblatex}
\usepackage{acronym}
\usepackage{graphicx}
\usepackage{hyperref}
\hypersetup{colorlinks=true, breaklinks=true, 
  urlcolor=blue, linkcolor=blue,anchorcolor=blue,citecolor=blue,
  hypertexnames=true, final=true, 
  pdfpagemode = UseNone, 
  pdfauthor = {},
  pdftitle = {},   
  pdfsubject = {},
  pdfkeywords = {}
}
\newcommand{\squeezeup}{\vspace{-2.5mm}}
\graphicspath{{img/} {figures/} {figures/circuits} {figures/intro_diagrams} {figures/results/} {./}}
\DeclareGraphicsExtensions{.pdf,.png,.jpg,.mps}

\IEEEdisplaynontitleabstractindextext

\DeclareSourcemap{
  \maps[datatype=bibtex]{
    \map[overwrite=true]{
      \step[fieldset=url, null]
      \step[fieldset=doi, null]
      \step[fieldset=eprint, null]
       \step[fieldset=month, null]  
    }
  }
}

\acrodef{EEG}[EEG]{Electroencephalography}
\acrodef{EMG}[EMG]{Electromyography}
\acrodef{HFO}[HFO]{High Frequency Oscillation}
\acrodef{ECG}[ECG]{Electrocardiography}
\acrodef{SNN}[SNN]{spiking neural network}
\acrodef{DPI}[DPI]{Differential Pair Integrator}
\acrodef{ADM}[ADM]{Asynchronous Delta Modulation}
\acrodef{RMSE}[RMSE]{Root Mean Square Error}
\acrodef{E-I}[E-I]{excitatory-inhibitory} 
\acrodef{SVM}[SVM]{Support Vector Machine}

\begin{document}
\title{
Feed-forward and recurrent inhibition for compressing and classifying high dynamic range biosignals in spiking neural network architectures
}

\author{
\IEEEauthorblockN{Rachel Sava, Elisa Donati, Giacomo Indiveri}
\IEEEauthorblockA{Institute of Neuroinformatics, University of Zurich and ETH Zurich Switzerland}
\thanks{This work has been submitted to the IEEE for possible publication. Copyright may be transferred without notice, after which this version may no longer be accessible.}
\thanks{This work was partially supported by the European Research Council (ERC) under grant agreement No 724295 (``NeuroAgents'')}


}

\maketitle

\begin{abstract}
  Neuromorphic processors that implement Spiking Neural Networks (SNNs) using mixed-signal analog/digital circuits represent a promising technology for closed-loop real-time processing of biosignals. As in biology, to minimize power consumption, the silicon neurons' circuits are configured to fire with a limited dynamic range and with maximum firing rates restricted to a few tens or hundreds of Herz.
  However, biosignals can have a very large dynamic range, so encoding them into spikes without saturating the neuron outputs represents an open challenge.
  In this work, we present a biologically-inspired strategy for compressing this high-dynamic range in SNN architectures, using three adaptation mechanisms ubiquitous in the brain: spike-frequency adaptation at the single neuron level, feed-forward inhibitory connections from neurons belonging to the input layer, and Excitatory-Inhibitory (E-I) balance via recurrent inhibition among neurons in the output layer.
  We apply this strategy to input biosignals encoded using both an asynchronous delta modulation method and an energy-based pulse-frequency modulation method.
  We validate this approach \textit{in silico}, simulating a simple network applied to a gesture classification task from surface EMG recordings. 
\end{abstract}


\section{Introduction}


Closed-loop systems allow personalized healthcare devices to detect unexpected changes continuously, and to provide feedback stimulation or control signals to maintain a desired state. Thanks to their ultra-low power consumption, biologically plausible time constants, and inherent parallelism, neuromorphic technologies provide ideal characteristics for real-time closed-loop interactions with the biological system.
Neuromorphic solutions enable always-on continuous monitoring of physiological parameters in a wide range of wearable applications, such as \ac{ECG} anomaly detection~\cite{Bauer_etal19, Das_etal18a, gerber2022neuromorphic}, \ac{HFO} detection~\cite{sharifshazileh2021electronic, burelo2022neuromorphic}, \ac{EMG} decoding for prosthetic control~\cite{Ma_etal20,Donati_etal19, vitale2022neuromorphic}, and nervous system interfacing ~\cite{donati2023neuromorphic}.

However, processing biosignals can be challenging because they have a high dynamic range in their frequency components and amplitude, often covering over 3 orders of magnitude. For example, in \ac{ECG}, the amplitude signal includes both large peaks from the louder backdrop of the heart beating and small fluctuations related to specific conditions~\cite{gerber2022neuromorphic}.
In the frequency domain, different bands are often associated with different underlying physiological processes, which are superimposed at the sensor interface~\cite{Scarsella}. For example, \ac{ECG} and \ac{EMG} signals capture both low-frequency drift in heart rate variability or muscular posture (0 - 5\, Hz) and rapid high-frequency electrical events ($ > $1\,kHz)~\cite{Martinek}.

\begin{figure}
  \centering
  \includegraphics[width=0.47\textwidth]{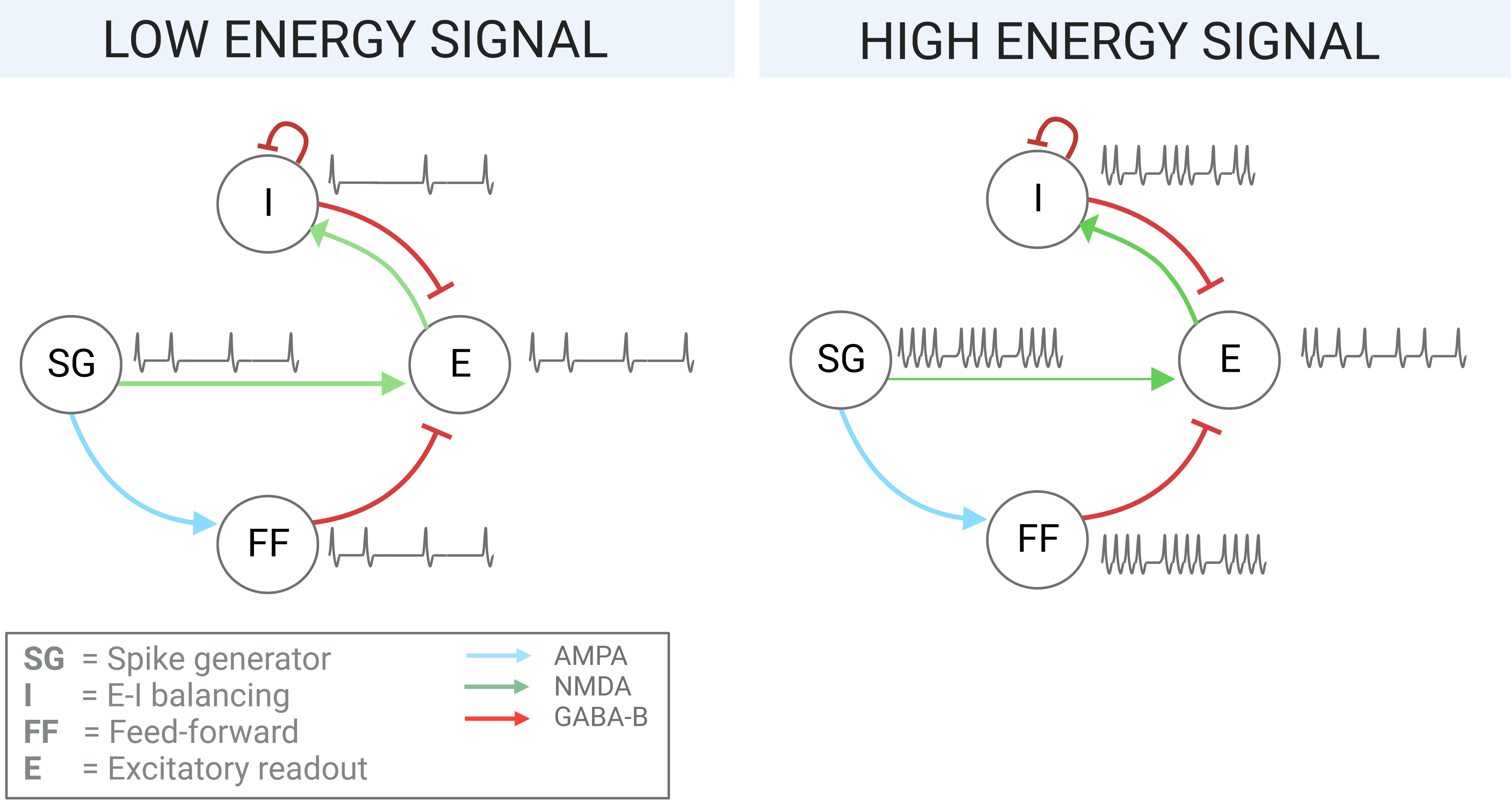}
  \caption{SNN-based frequency balancing mechanism, comprising a feed-forward inhibitory (FF) population and an excitatory-inhibitory balanced subnetwork (I, E). }
  \label{fig:EIFF}
\end{figure}

As one core function of the brain is to decipher real-world events from perceptual information, it is often required to encode and process exceptionally high dynamic range signals (for example, auditory neurons encode noise from infrasound ($ < $20\,Hz) to ultrasound ($ > $20\,kHz))~\cite{Wen}.
To discriminate stimuli over this large frequency range, the brain employs various structural, biochemical, and cortical mechanisms of adaptation.
Among them are \ac{E-I} balanced neural networks, which develop a near-constant ratio between inhibitory and excitatory synaptic currents in each cell. This renders them able to suppress excessive firing for high-frequency inputs and facilitate activation for low-frequency inputs~\cite{Sohal} (see Fig.~\ref{fig:EIFF}).
The resulting network activity is sparse and decorrelated, which minimizes information redundancy and leads to a more energy-efficient encoding~\cite{Zhou}.
A second neural adaptation mechanism is feed-forward inhibitory connections, which sharpen the temporal precision of the target neurons by narrowing the window for suprathreshold summation of excitatory inputs by hyperpolarizing the membrane potential shortly after the onset of excitation.
This enhances signal discrimination and prevents saturation in neural circuits~\cite{Roberts}.
Finally, at the cellular level, neurons exhibit spike-frequency adaptation as an intrinsic mechanism to prevent premature saturation: Calcium ion accumulation during prolonged or repetitive output spikes activates potassium channels, generating a hyperpolarizing current that reduces the neuron's excitability and firing rate. This mechanism preserves the dynamic range of neuronal responses and ensures responsiveness to changing input signals~\cite{Ha}, and has been shown to be equivalent Sigma-Delta encoding schemes~\cite{Nair_Indiveri19} which enable accurate and efficient time-domain classification~\cite{yin2021accurate}. 

In this paper, we present a \ac{SNN} architecture implemented to address frequency mismatch for biosignal processing and to facilitate a wide range of classification tasks. Through exploiting inhibitory adaptation mechanisms ubiquitous in the brain, this ultra-compact network is maintained in effective operating range, independently of the input dynamic range. To validate the behavior of the network, we show its implementation for the processing of \ac{EMG} signal for hand gesture classification. We are targeting \ac{EMG} signals due to their frequency (a few Hz – 1 kHz) and amplitude (a few V - 10 mV) ranges. In addition, for the first time, we show a comparison for signal-to-spike conversion required to feed the analog inputs to the \ac{SNN}. We present an asynchronous delta modulation method (mainly sensitive to the changes in the input signals)~\cite{sharifshazileh2021electronic} and an energy-based method (sensitive also to the DC level of the input signals). 

\section{Network implementation}
\label{sec:network}

The network proposed has been designed to be compatible with a wide range of mixed-signal neuromorphic processing chips. However, to assess its biological plausibility and robustness for deployment on-chip, we first simulated it using neural and synaptic models based on the Dynamic Neuromorphic Asynchronous Processor (DYNAP-SE)~\cite{Moradi_etal18}. 

\subsection{Neuromorphic behavioral simulator}
The \ac{SNN} was simulated using the spiking simulator Brian2~\cite{brian2} using custom equations and parameters that faithfully model the transistor and circuit properties present in the DYNAP-SE chip.
In particular, neural and synaptic dynamics were modeled based on the \ac{DPI} circuit, which implements a first-order log-domain filter with non-linear properties that are useful to reproduce short-term plasticity effects~\cite{Bartolozzi_Indiveri07a}.
The DYNAP-SE neurons are connected on-chip by synapses of classes NMDA (slow excitatory), AMPA (fast excitatory), GABA-A (fast inhibitory), and GABA-B (slow inhibitory). The neurons are divided into four cores, and the properties of the neurons of each core as well as each synapse class are tunable via the selection of relevant circuit parameters (e.g. leak and gain currents). Arbitrary subsets of the 256 available neurons may be connected with synapses of pre-specified weight.

\subsection{Signal pre-processing}

An \ac{EMG} dataset was previously collected for 3 gestures from the Roshambo game (rock, paper, scissors) using the Myo armband by Thalmic Labs Inc (8 forearm sensors sampled at 200 Hz) from 10 able subjects over 3 sessions each containing 5 repetitions of $2~seconds$ of gesture~\cite{elisa:dataset}. Between each session, a relaxing phase of 1$~second$ was introduced to remove any residual muscular activation. 

The recordings were then segmented into $200$ $~ms$ windows -- an accepted latency in prosthetic control~\cite{Smith} -- and shuffled to generate the training and test datasets. To reduce the impact of label bleeding (where muscular activity persists in the neighboring rest period), a function was employed to exclude windows that more resemble the adjacent state than the current. Oversampling was then employed to balance the class occurrences. 

To serve as input to the \ac{SNN}, raw voltage traces were converted into the event-based domain. During the conversion into spikes, it is crucial to maintain the original signal information. With this aim, for the first time, we applied and compared two bioinspired approaches. 

\subsection{Asynchronous Delta Modulation}
The high-frequency \ac{ADM} algorithm, based on a delta modulator circuit~\cite{Corradi_Indiveri15}, was applied to transform the continuous \ac{EMG} signals into two digital pulse outputs ('UP' and 'DOWN') according to the signal derivative (positive and negative, respectively) (Fig.~\ref{fig:converted}). The parameters were optimized by a grid search to minimize the \ac{RMSE} between the original signal and a reconstruction from the \ac{ADM} spike trains. The final \ac{ADM} parameters are a threshold of 0.8 --corresponding to the sensitivity for the spike conversion --, a  refractory period of 10 $\mu$ s -- a period of system inactivity after the generation of a spike --, and an interpolation factor of 3500 -- a factor required to increase the signal resolution. The resulting spike trains contain frequency elements from 0 - 8 kHz.


\subsection{Energy-based Pulse Frequency Modulation}
\begin{figure} [!ht]
            \centering
            \includegraphics[width=0.47\textwidth]{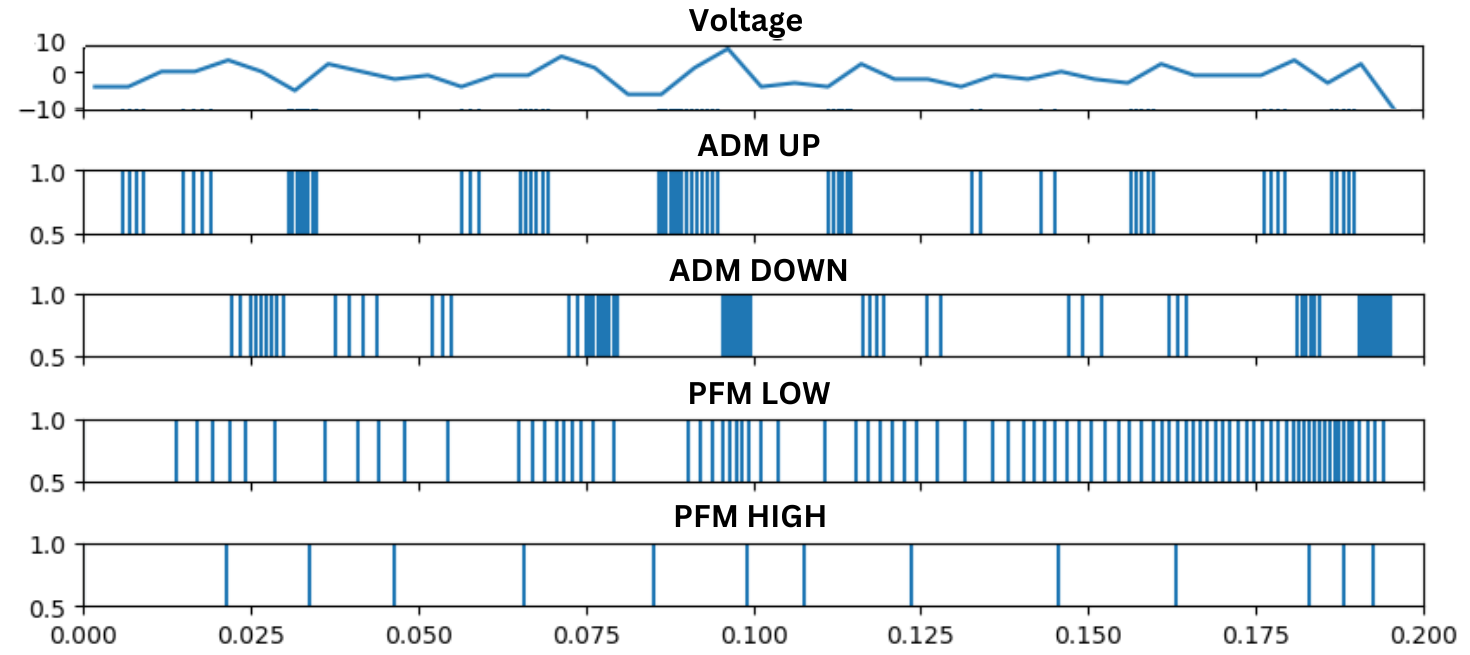}
            \caption{Input signal and the its conversion into spike trains by \ac{ADM} (up and down channels) and energy-based method on frequency bands (low and high bands). }
            \label{fig:converted}
    \end{figure}

An energy-based approach was devised as an alternative mechanism for spike conversion. The raw signal was filtered into 2 frequency bands split equally between 0 and 100\,Hz, full-wave rectified, and injected as a time-varying current into a simple integrate-and-fire Brian2 neuron to achieve pulse-frequency modulation  (Fig.~\ref{fig:converted}). To ensure that the resulting firing rate is proportional to the energy of the original signal, the input values were scaled to best cover the linear range of the neuron's injected current (0 - 8 $~nA$) vs firing rate (0 - 4 kHz) curve. Since - due to high-amplitude outliers - a direct compression of the original range to the target results in most values occupying the lowest fifth of the linear region, for each subject and session a unique scaling range was calculated to best redistribute the mass of the amplitude distribution, and employed across all channels to maintain the relative amplitude differences between electrodes. The resulting spike trains contain frequency elements from 0 - 3 kHz.

\subsection{Architecture and behavior tuning}

\begin{figure}
            \centering
            \includegraphics[width=0.47\textwidth]{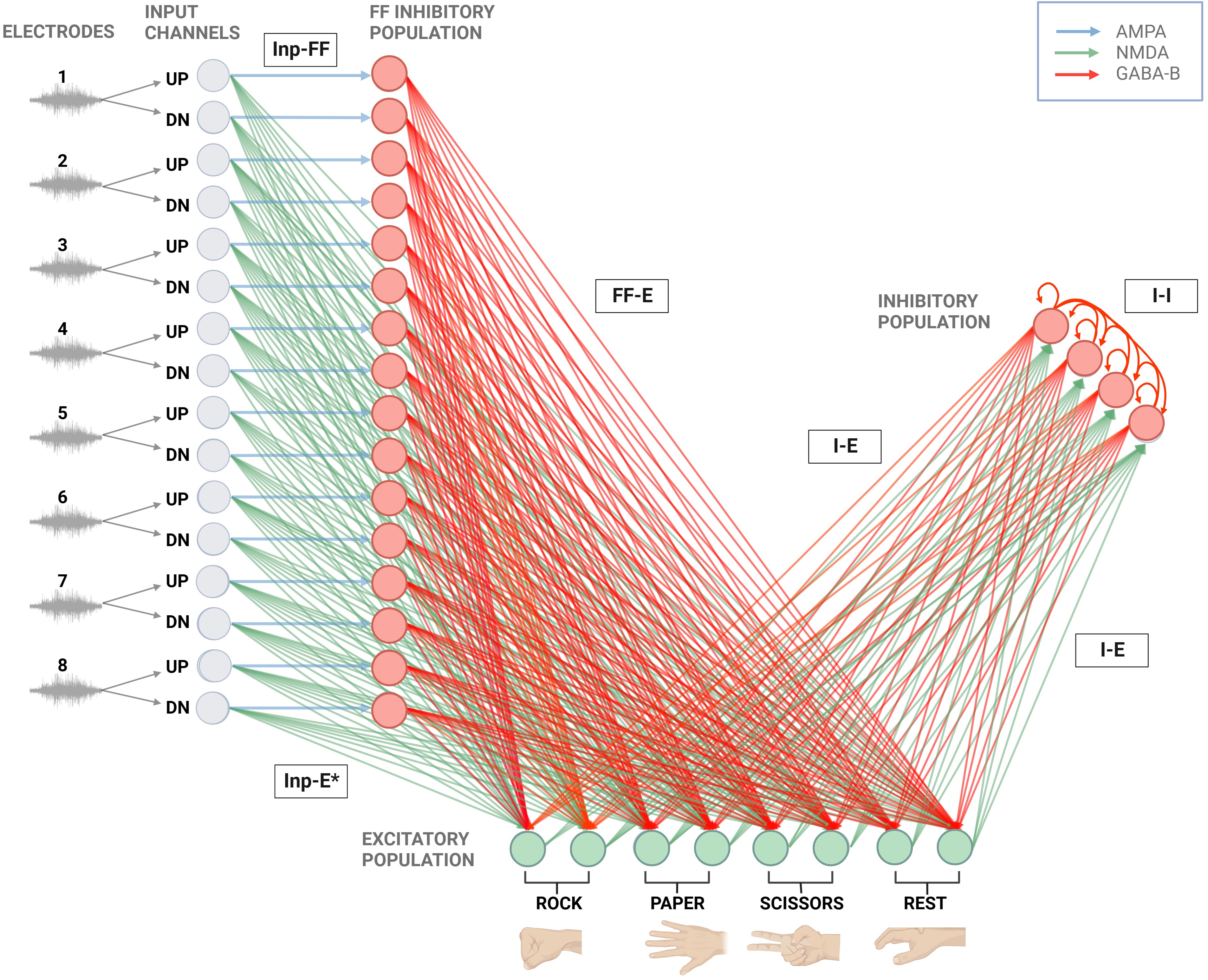}
            \caption{Architecture of the full SNN.}
            \label{fig:fullnet}
    \end{figure}

The \ac{SNN} consists of an input population of 16 neurons, an excitatory population of 8 neurons which also serve as the classification output layer, a feed-forward inhibitory population of 16 neurons, and a recurrently-connected inhibitory population of 4 neurons (Fig.~\ref{fig:fullnet}). Together, the inhibitory populations facilitate network-level spike-frequency adaptation (Fig.~\ref{fig:EIFF}). In response to low energy signal, patterns are propagated to the excitatory readout layer, the feed-forward population fires sparsely, and the E-I subnetwork converges on a low firing rate via mutual recurrent inhibition. For high energy signals, the spike generator strongly activates the feed-forward population which drives down activity in the readout layer. This layer also promotes its own inhibition via excitation of the E-I subnetwork, which progressively drives down its own activity to a steady state via self-inhibition. These complementary mechanisms maintain the readout layer in the ideal operating range.

\begin{figure} [!ht]
            \centering
            \includegraphics[width=0.47\textwidth]{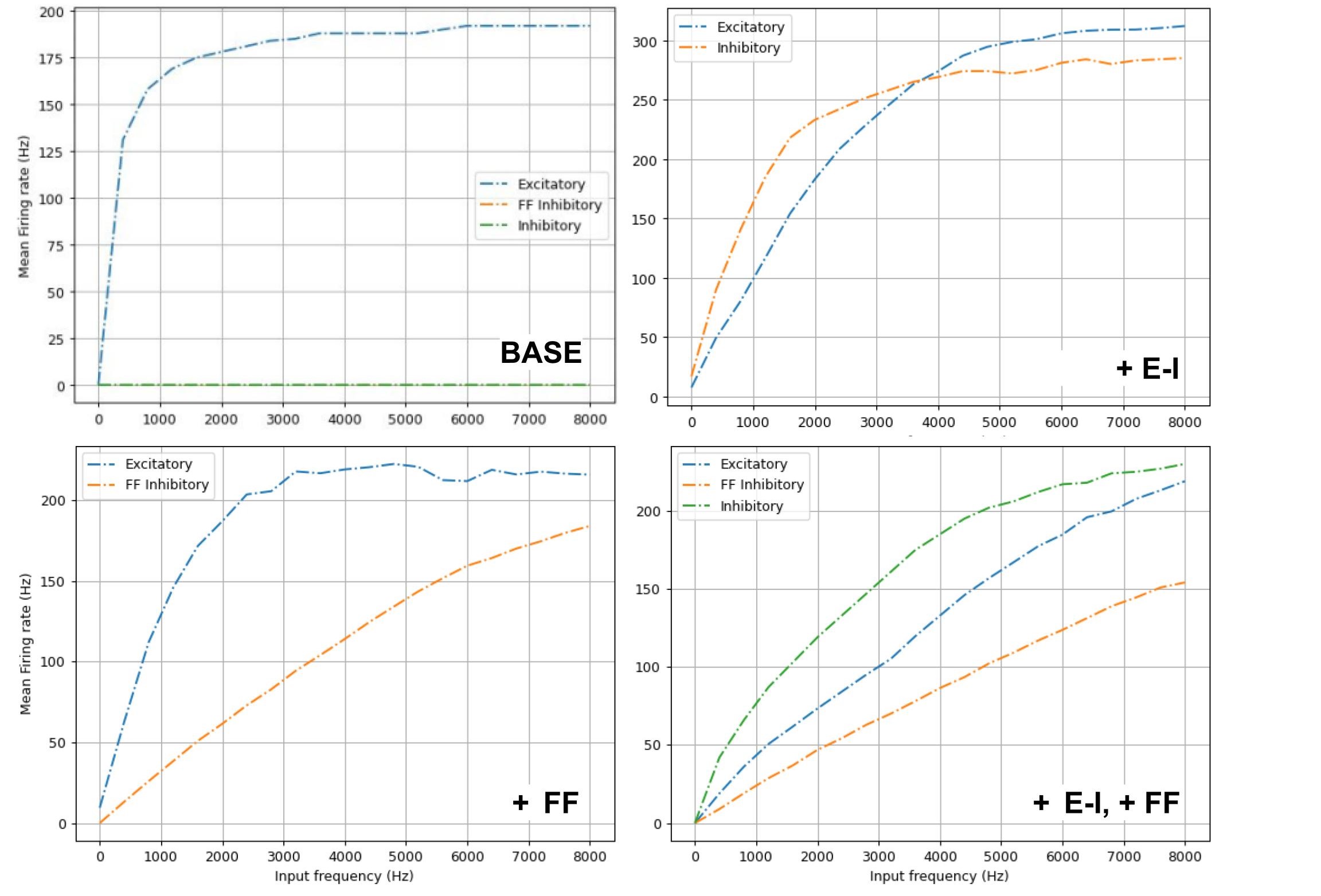}
            \caption{Input-output curves for the full network with progressive addition of inhibitory neuron populations.}
            \label{fig:inout}
    \end{figure}

The parameters of all populations (Table~\ref{tab:properties}) were progressively optimized via grid search to achieve linearity in the input-output curves of the classification layer neurons to input patterns ranging from 0 to 8\,kHz (Fig.~\ref{fig:inout}). This range was selected to fit the higher maximum spike train firing rate resulting from the ADM conversion method. 

Poisson spike trains at each frequency were employed in the network tuning to generate input-output curves (Fig.~\ref{fig:inout}). 
To account for variability and device mismatch in future hardware implementations of the network proposed, we added external noise by sending additional 40\,Hz Poisson spike trains through AMPA and GABA-B synapses to each neuron, combining fast and slow excitatory and inhibitory modulating currents.

\begin{table}
\centering
\setlength{\tabcolsep}{3.5pt}
\renewcommand{\arraystretch}{1.3}
\begin{tabular}{|>{\normalfont}l|l|}
\hline
\multicolumn{2}{|c|}{\textbf{Synapse}} \\
\hline
Inp-FF & 0.5 \\
FF-E & 3.8 \\
Inp-E & trained \\
I-E & 3.0 \\
E-I & 1.7 \\
I-I & 0.5 \\
\textnormal{Leak current} & 3 \textit{pA} \\
\textnormal{Gain current} & 12 \textit{pA} \\
AMPA $I_\tau$ & 10 \textit{pA} \\
GABA-B $I_\tau$ & 10 \textit{pA} \\
NMDA $I_\tau$ & 5 \textit{pA} \\
\hline
\end{tabular}
\hspace{8pt}
\begin{tabular}{|>{\itshape}l|l|}
\hline
\multicolumn{2}{|c|}{\textbf{Neuron}} \\
\hline
\textnormal{Membrane capacitance} & 2 \textit{pF} \\
\textnormal{Refractory period (E)} & 3 \textit{ms} \\
\textnormal{Refractory period (I)} & 1 \textit{ms} \\
\textnormal{Refractory period (FF)} & 1 \textit{ms} \\
\textnormal{Leakage current} & 5 \textit{pA} \\
\textnormal{Spiking threshold current} & 2000 \textit{pA} \\
\textnormal{Reset current} & 1.2 \textit{pA} \\
\textnormal{Constant current injection} & 1 \textit{pA} \\
\textnormal{Gain current} & 5 \textit{pA} \\
\textnormal{Adaptation} $I_\tau$ & 0.04 \textit{pA} \\
\textnormal{Adaptation gain} & 1.5 \\
\hline
\end{tabular}
\caption{Parameters of Adaptive Integrate–and–Fire (AIF) model neurons and synapses\protect~\cite{Bartolozzi_Indiveri07a}. }
\label{tab:properties}
\end{table}

\squeezeup
 \subsection{Training and testing}
The \ac{EMG} dataset was split into $200$ ms windows and converted into two spike trains for each of the 8 electrodes (UP/DOWN or High/Low). 
The dataset was divided in a 80\%-20\% train-test split. After the training, the weight connection between the inputs and the excitatory population, \textit{Input-E}, were frozen, and the output neuron pair with the highest firing rate over each test window was taken as the prediction of the network. 

We applied a supervised learning method to update the Input-E weights according to the delta learning rule: 
$ \Delta w_{ji} = \alpha (T_{j}-y_{j})x_{i} $. To accomplish this, two values were added to the DYNAP-SE neuron equations: a teacher value ($T$) (0.1 for the correct class, 0 for all others) and the instantaneous firing rate ($y$) (approximated by an exponential kernel ~\cite{Taro} whose decay is defined by $\tau_x\frac{d x_{trace}}{dt}=-xtrace$, where $\tau_x$ is 50 ms). A low learning rate of 5e-4 was selected. 

\squeezeup
\section{Results}
\subsection{Network behavior}
\begin{figure}
            \centering
            \includegraphics[width=0.46\textwidth]{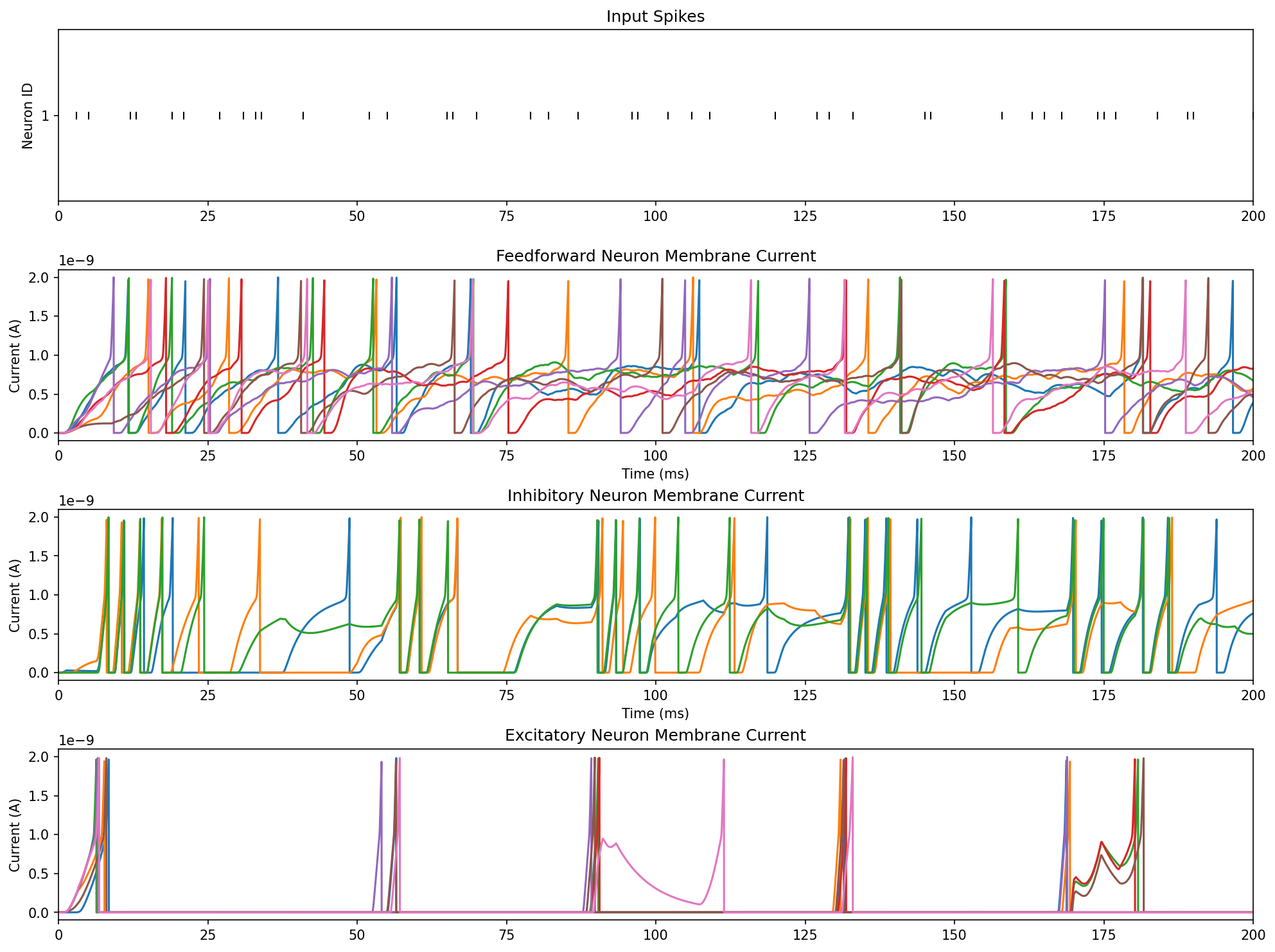}
            \caption{Representative neuron population behaviors in response to 1000 Hz Poisson input.  }
            \label{fig:finaladap}
        \end{figure}

Figure~\ref{fig:finaladap} highlights the desired behaviors of each neural group in the final network: the \textit{FF} population fires in step with the spike generator, maintaining its target layer in reasonable operating range by coupling the strength of excitatory input to the level of inhibition. The \textit{recurrent-I} population fires more dominantly than the readout layer, maintaining the inhibitory-dominated E-I network. These result in the \textit{E} population firing sparsely. 

\subsection{Classification accuracy}
Across all train-test cycles for a subset of subjects with the highest previously reported classification accuracies (2, 3, and 4) \cite{elisa:dataset}, the mean accuracy (three-fold validated) of the \ac{SNN} was compared to a \ac{SVM} trained on either the data after root mean squared (window = 10 samples, optimized via grid search) which yielded the highest accuracy, or - as a better comparison for the spike conversion methods - the raw electrode traces (kernel = RBF, c=10, optimized via grid search).  

\begin{figure}
            \centering
            \includegraphics[width=0.47\textwidth]{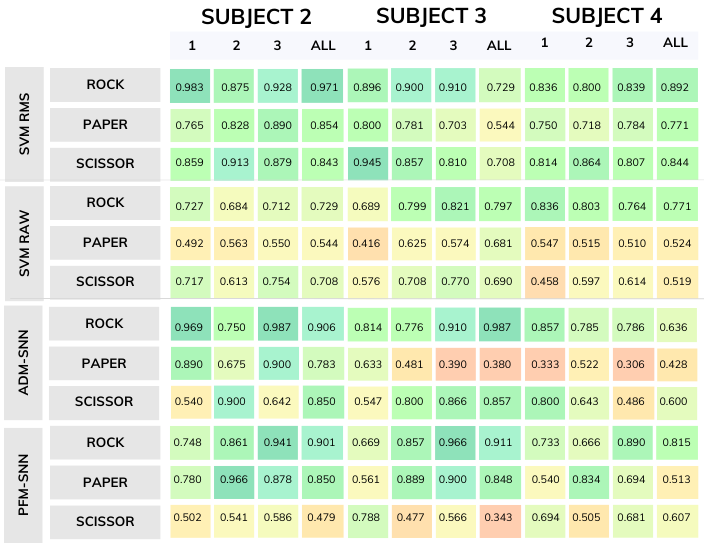}
            \caption{Mean accuracy percentage across subject and session for SVM trained on RMS data (window = 10), SVM trained on raw traces, ADM conversion + SNN, and PFM conversion + SNN.\protect\footnotemark  }
            \label{fig:slide}
        \end{figure}
\footnotetext{Colour reflects accuracy percentage. For each subject, training and testing were performed per session (1, 2, 3) and on pooled sessions (1+2+3), 'ALL'.}


To ascertain whether the network's behavior is generalizable across individuals, the ADM-SNN was trained and tested on all data from all subjects. The results of this training showed comparable accuracy to the individual sessions (65.04\% ±3.89\%, n=3), suggesting the network's robust ability to abstract temporal features despite varied baseline activity levels.

\subsection{Base-to-full analysis}
To quantify the contributions of each network element, the network was progressively constructed and tested on identical data (Subject 3.3). 
The plastic synaptic weights in networks without both inhibitory populations tend to lower values over the course of training (Fig. \ref{fig:slide}), due to the saturation of the excitatory layer in the absence of frequency-modulating mechanisms. 

\begin{figure}
            \centering
            \includegraphics[width=0.47\textwidth]{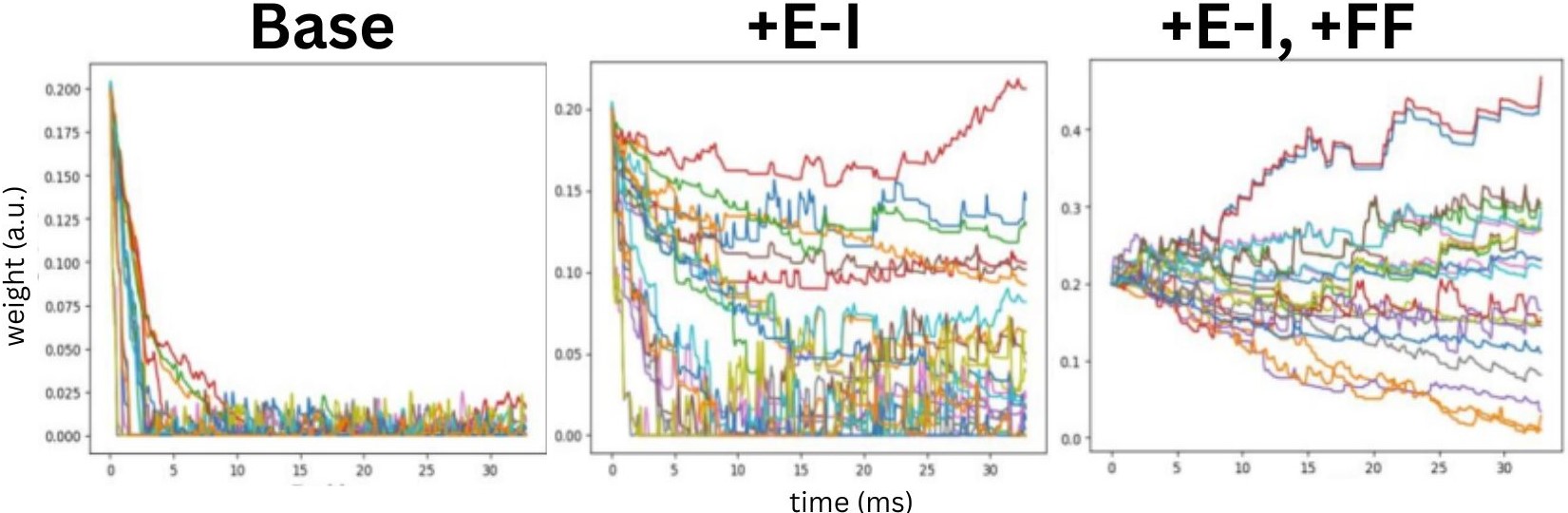}
            \caption{Weight evolution upon progressive network additions}
            \label{fig:slide}
        \end{figure}

\textit{FF} inhibitory neurons appear to most significantly reduce saturation in the output layer, and thus in the prediction accuracy (Table~\ref{tab:accuracy}), suggesting that the greatest hindrance to \ac{SNN} classification is an inability to capture input patterns at higher firing frequencies. Excitatory-inhibitory mechanisms also serve this aim, but prove insufficient on their own. 
\begin{table}[h!]
\centering
\begin{tabular}{lllll}
\hline
Base & +spike adapt   & +E-I  & +FF & Full   \\ \hline
27.3\%       & 47.1\% & 73.3\% & 82.5\%  & 89.0\% \\
\end{tabular}
\caption{Classification accuracy for progressive network configurations (additions are relative to Base).}
\label{tab:accuracy}
\end{table}

\section{Conclusions}
We proposed an \ac{SNN} architecture that offers a low-latency solution to the processing of high dynamic range biosignals in neuromorphic networks. While spike conversion methods yield input spike trains with frequency components that would otherwise saturate the firing of the DYNAP-SE neurons in plain feed-forward networks, the \acp{SNN} architecture proposed proved capable of compensating for the increased dynamic range of the input signal in simple class-based classification tasks, making neuromorphic chips prime candidates for lightweight, long-lasting on-board electronics for bio-interfacing controllers. 





\printbibliography


\end{document}